# CryoCMOS RF multiplexer for superconducting qubit control, readout and flux biasing at millikelvin temperatures with picowatt power consumption

L. Fallik[1,2], S. Balamurali[1], A. Caglar[1], R. Acharya[1], J. Van Damme[1], Ts. Ivanov[1], S. Massar[1], R. Asanovski[1], A. M. Vadiraj[1], M. Mongillo[1], J. Craninckx[1], A. Grill[1], D Wan[1], A. Potočnik[1]*, K. De Greve[1,3]

[1]Imec, Kapeldreef 75, Leuven, 3001, Belgium,
[2]Department of Electrical Engineering (ESAT), KU Leuven, Leuven, 3001, Belgium
[3]Proximus Chair in Quantum Science and Technology, department of Electrical Engineering (ESAT), KU Leuven, 3001 Belgium

*Corresponding author: anton.potocnik@imec.be

Large-scale cryogenic quantum systems are constrained by an input–output bottleneck between room-temperature electronics and millikelvin stages, particularly in superconducting qubit platforms. This bottleneck is most acute for output lines, where bulky and expensive microwave components limit scalability. A promising approach for scalable characterization and testing is to perform signal multiplexing directly at the qubit plane. We demonstrate a cryogenic CMOS (cryoCMOS) RF multiplexer operating at 10 millikelvin with record-low static power consumption of 200 pW. The device provides < 2 dB insertion loss and > 30 dB isolation across DC-8 GHz. Direct connection to transmon qubits marginally affects coherence times in the range of 100 μs, enabling multiplexing of readout, flux and, in principle, XY drive lines. This work introduces cryoCMOS multiplexers as valuable tools for scalable, high-throughput cryogenic characterization and testing, and advances co-integrated quantum–classical control for future large-scale quantum processors.

## I. Introduction

Quantum computers hold the promise of revolutionizing information processing by performing computations beyond the current capabilities of classical computers – offering a range of potential applications in cryptography, quantum physics and chemistry and materials science among others[1–5]. To account for unavoidable, physical errors in the qubits and in their control, quantum error correction is being widely pursued, where a logical qubit is encoded in a large number of physical qubits[6]. As a consequence, achieving error-corrected, fault-tolerant quantum computation will require the integration of millions of interacting qubits[7], far surpassing current state-of-the-art systems containing only a few hundred[8]. A key bottleneck in the operation and characterization of large-scale quantum processors lies in their control and readout infrastructure[9]. Each qubit typically requires multiple DC and/or RF control lines inside a dilution refrigerator, delivering signals from room-temperature electronics to qubits operating at millikelvin temperatures.

Cryogenic CMOS electronics (cryoCMOS) have the potential to alleviate this bottleneck, by enabling signal generation[10–12] and detection[13,14] directly at cryogenic temperatures. However, the limited cooling power at millikelvin temperatures (~20 μW)[15], severely constrains the integration of active components, and current cryoCMOS implementations often exceed this power limit. Significant research and development in cryoCMOS is still required to drastically reduce both static and dynamic power consumption[16,17].

A promising near-term application of cryoCMOS at millikelvin temperatures is signal multiplexing. It is, however, important to note that multiplexing, most notably time-division multiplexing (TDM), does not provide a scalable solution to the wiring bottleneck for operating quantum processors with many qubits simultaneously. Even in the ideal case, TDM can only achieve a theoretical reduction in the number of signal lines by a factor of 0.677[18], representing an incremental rather than transformative improvement. Its practical relevance therefore lies primarily in large-scale quantum device testing and characterization[15], where sequential device measurements are acceptable (Figure 1a). This is feasible, provided that the multiplexers exhibit appropriate microwave properties, such as low insertion loss at microwave frequencies to enable the multiplexing of the extremely weak signals used to probe quantum devices, low crosstalk, low 1/f noise, and most importantly, sufficiently low broadband thermal noise so as not to affect the qubits' coherence times. These requirements are particularly stringent for the control and readout of noise-sensitive superconducting qubits[19,20].

Time-division RF signal multiplexing at millikelvin temperatures has been demonstrated using a range of technologies, including BiCMOS[21], III-V two-dimensional electron gas (2DEG) with superconducting Nb metalization[22] and Josephson junction-based field effect transistors (JoFET)[23]. While these approaches establish the feasibility of cryogenic TDM, their compatibility with qubit operation has yet to been verified. The Single Flux Quantum (SFQ) technology has also been proposed for TDM[24]; however, its applicability is restricted to single flux quantum pulses and does not extend to broadband RF signals. More recently, MEMS-based switches have emerged as promising candidates for millikelvin multiplexing[25], offering near-zero static power consumption, excellent microwave isolation (> 40 dB) and insertion loss (< 1 dB). At the same time, their use near qubits raises concerns related to the release of Cooper-pair–breaking phonons during mechanical actuation, potentially shortening qubit relaxation times, as well as long-term reliability issues associated with charging-induced stiction[26].

Alternatively, the cryoCMOS technology offers a solid-state, fast, reliable and potentially low-power solution that benefits from decades of industrial development, resulting in a mature technology that can be manufactured at scale. CryoCMOS multiplexers have already been demonstrated for both semiconducting (spin) qubits[16,27,28] and superconducting qubits[20,29]. The most recent demonstration of cryoCMOS signal multiplexing of superconducting qubit input lines successfully showed qubit driving through the multiplexer, but was limited by its relatively high static power dissipation and thermal noise, which notably shortened qubit dephasing times and required additional attenuation between the multiplexer and the qubits[20].

In this work, we present a novel cryoCMOS single-pole four-throw (SP4T) RF multiplexer with a record-low static power consumption of 200 pW, representing a 3000-fold improvement compared to our previously reported state-of-the-art implementation[20]. The device supports signal multiplexing for all superconducting qubit processor signal lines, including microwave input and output lines for qubit driving and readout, low-frequency flux-biasing (Figure 1a), and in principle also the multiplexing of XY drive lines (not shown in this work). When connected to the common feedline used to excite and measure the qubits' state, the multiplexer has a negligible effect on energy relaxation times ($T_1$), and causes only a small reduction in coherence times ($T_{2e}$) through photon shot noise, corresponding to an inferred added photon flux of at most ~0.02 photons/Hz/s (~$10^{-26}$ W/Hz at 6 GHz). When the multiplexer is connected to the flux-bias line, no detrimental effects of the 1/f noise can be observed on the qubit's coherence times. This performance was achieved by employing a custom-made multiplexer design based on a commercial Fully Depleted Silicon-On-Insulator (FDSOI) technology and specifically targeting near-zero power consumption at 10 mK.

## II. Device description

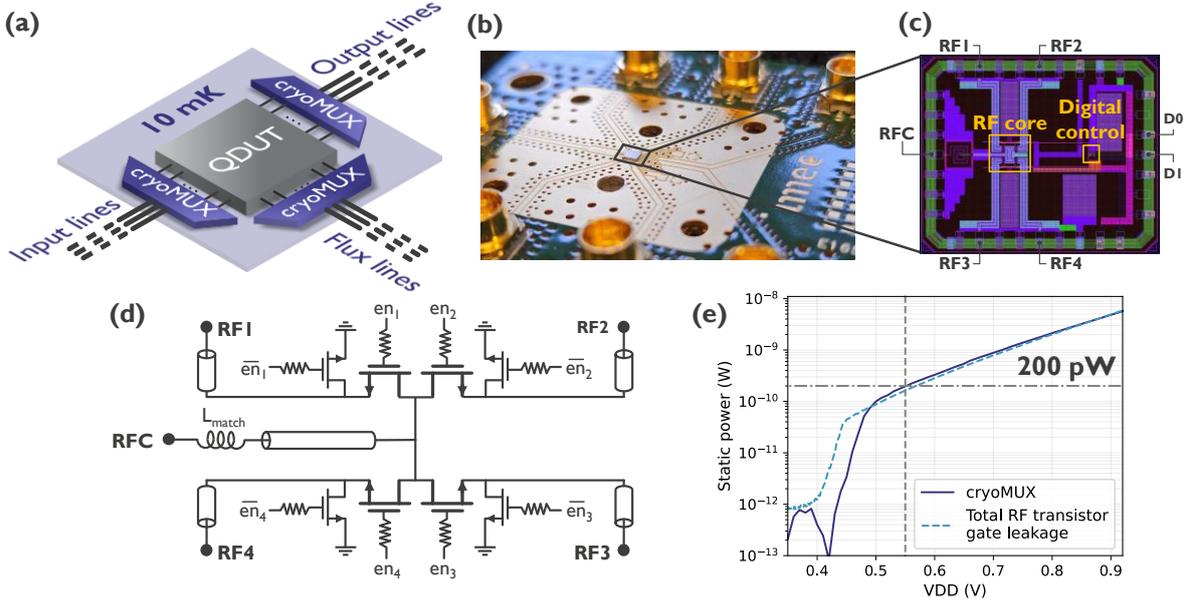

*Figure 1: CryoCMOS multiplexers for scalable superconducting qubit control, readout and flux biasing operating at 10mK. (a) Schematic of a scalable quantum device (QDUT – Quantum Device Under Test) characterization setup using cryoCMOS multiplexers (cryoMUX) for input, output and flux-bias lines. (b) Photograph and (c) layout of the multiplexer die. (d) Circuit schematic of the RF circuit. The digital signals ($en_i$) are supplied by the digital control circuit (not shown). Series RF transistors are depicted larger than shunt transistors to reflect their relatively larger size (not to scale). (e) Static power consumption as a function of the supply voltage (VDD). A consumption of 200 pW was measured at a supply voltage of 0.55 V (grey horizontal dot-dashed line), the lowest voltage where the multiplexer is fully operational. The light-blue dashed line represents the cumulative gate leakage current of all transistors in the RF core that are active simultaneously (one series and three shunt transistors).*

A Single Pole Four Throw (SP4T) cryoCMOS RF multiplexer implemented using four sets of series-shunt switch arms, is designed and fabricated in a commercial 22 nm FDSOI technology (Figure 1 b-d). Parallel logic control is implemented with two digital input lines (D0 and D1) to set the signal path through the multiplexer from the common port (RFC) to one of the four output ports (RF1-4). The circuit is optimized for ultra-low static power consumption at cryogenic temperatures and for broadband RF performance in the DC-10 GHz range. This includes the use of dedicated clamp transistors in the electrostatic discharge (ESD) protection cells, low-$V_T$ (*lvt*) type transistors for the digital control circuits, and super-low-$V_T$ (*slvt*) type transistors for the RF switch transistors. To maintain a low insertion loss and high isolation, relatively large series RF transistors are employed, with a total area of 1.28 µm² and shunt pull-down transistors with an area of 0.32 µm².

Measurement of the static power consumption as a function of supply voltage (VDD) at 10 mK shows a turn-on VDD of ~0.5 V, which is the threshold voltage of the *lvt* transistors in the digital control circuit at 10 mK. This threshold voltage increases by approximately 0.2 V at 10 mK relative to its room-temperature value[30]. Above the turn-on point, the static power consumption increases exponentially as function of VDD, indicating that the power consumption is dominated by a leakage current tunnelling through the thin RF transistor gate oxides[31].

When operated at VDD = 0.55 V, the multiplexer exhibits a static power dissipation of only 200 pW, attributed to the overall low leakage in the FDSOI technology and to the design optimizations discussed above. Based on I-V measurements of an isolated transistor at 10 mK (dashed line in Figure 1e), most of the static power consumption is identified as originating from the gate leakage of the active series and pull-down RF transistors. The static power dissipation of the included ESD cells is below the measurement limit (< 100 fW, see Supplementary Figure 2) and therefore negligible.

Although dynamic power consumption is less important for device testing and characterization applications where switching times are expected on the order of minutes to hours, it is nevertheless critical for future, more advanced cryoCMOS applications operating at high speeds at millikelvin temperatures. When rapidly toggling D0, a dynamic power dissipation of 108 fJ per switching event is measured at a VDD of 0.55 V at 10mK, which is comparable to previous implementations[20] (see Supplementary Figure 3). Dynamic power dissipation originates from the charging and discharging of the RF transistors' gate capacitors and the parasitic capacitance in the long interconnects between digital and RF circuits, as confirmed by circuit simulations and as expected for a CMOS implementation.

Microwave signal transmission through the multiplexer increases sharply at VDD = 0.48 V when the device turns on, followed by a slower increase until ~ 0.55 V after which the transmission saturates (Figure 2a). Based on these measurements, VDD = 0.55 V is chosen as the optimal operating point, as it provides the best trade-off between microwave transmission and the static power performance described above. At this operating point, the insertion loss remains below 3 dB up to 9.2 GHz (Figure 2b). The measured insertion loss is comparable to simulations performed at -40 °C, with the difference attributed to the temperature difference, variations between fabricated devices and simulation models, packaging parasitics, impedance mismatch and wire-bond inductance. Overall, the insertion loss is comparable to typical cable and microwave components used in quantum computing experimental setups and is therefore not expected to affect the performance of qubit readout.

Port-to-port isolation is characterized by measuring the transmission through the RF1 port while the multiplexer is programmed to route the signal to one of the other three ports (RF2-RF4), which are terminated with a 50 Ω load. The device exhibited excellent isolation, exceeding 30 dB up to 8 GHz (Figure 2c). The data is again found to be comparable to simulations with calibrated -40 °C models. The slight reduction in measured isolation when compared to simulations is most likely due to crosstalk between PCB traces, which are not included in the simulations. From a test and characterization perspective, a port-to-port isolation of at least 30 dB up to 8 GHz is considered sufficient for scalable superconducting quantum device characterization[20].

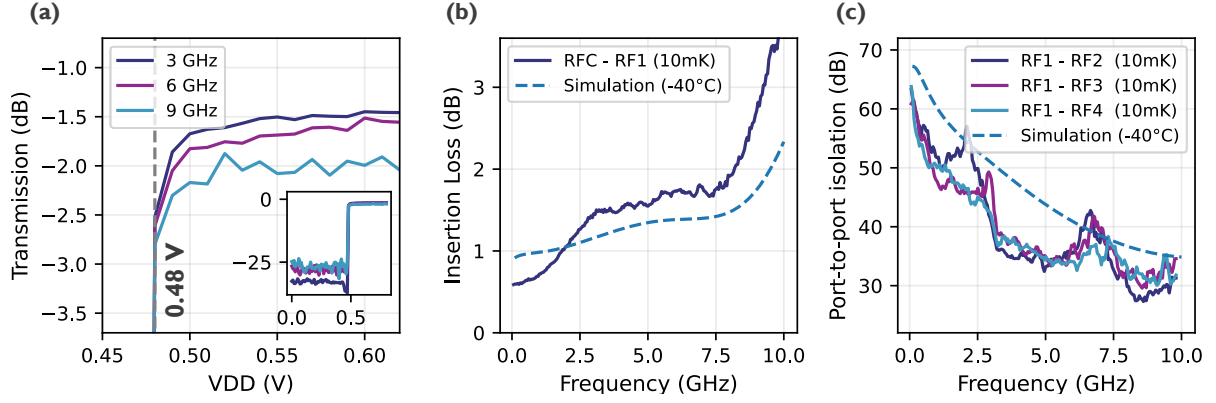

*Figure 2: cryoCMOS multiplexer RF performance at 10 mK. (a) Transmission as a function of supply voltage. The multiplexer turns on at 0.48 V (grey dashed line). (b) Transmission spectrum of the RF multiplexer, showing an insertion loss of around 1.5 dB for frequencies below 8 GHz at VDD = 0.55 V. (c) Port-to-port isolation, obtained by measuring transmission through port 1 while another port (2-4) was selected at VDD = 0.55 V. Dashed lines in (b) and (c) denote simulations performed at the lowest temperature available in the technology PDK (-40 °C), with VDD set to 0.55 V.*

## III. Superconducting qubit performance

The compatibility of the cryoCMOS multiplexer with superconducting qubits is evaluated on transmon qubits[19] fabricated at imec with a 300mm CMOS-compatible process[32]. The multiplexer is operated in three different configurations: (i) drive/readout input line multiplexing, (ii) readout output line multiplexing and (iii) flux-bias line multiplexing (Fig. 1a). For the first configuration, the impact of the multiplexer on qubit energy relaxation ($T_1$) and coherence times ($T_{2e}$) is measured and compared to reference values, as has been done in previous studies[20]. The qubit samples used for this experiment have a standard superconducting qubit circuit architecture, where each qubit is dispersively coupled to a coplanar waveguide (CPW) resonator, which is in turn coupled to a common coplanar-waveguide feedline used for carrying the input and output signals. XY qubit control (drive) signals are applied through the input line/feedline and reach the qubit through the readout resonator. Other designs might include dedicated drive lines for each qubit[19], but it is not the case here. On the qubit chip, one qubit has a dedicated standard flux-bias line for qubit frequency tuning (see Supplementary Figure 4).

The multiplexer's impact is first characterized when connected to the readout input line (Figure 3a). In this experiment, the feedline of a packaged transmon qubit chip is connected to the RF1 port of the multiplexer without any additional attenuation. The three remaining ports (RF2-4) are terminated with 50 Ω loads. A low-insertion-loss directional coupler (< 0.5 dB at RT) is incorporated into the signal chain to enable measurements when the multiplexer is off, allowing for the direct comparison of the qubits' coherence times to reference values within the same cooldown. In this configuration, any electromagnetic noise introduced by the multiplexer propagates through the feedline and creates a thermal photon population in the readout resonators[15]. These thermal excitations lead to a reduction of qubit dephasing times ($T_\phi$) due to photon shot noise[33] and consequently to a reduction of coherence times $T_{2e}$ (where $\frac{1}{T_{2e}} = \frac{1}{T_\phi} + \frac{1}{2T_1}$). The dephasing time limit due to the thermal photon shot noise in the readout resonator is given by[34]:

$$1/T_\phi = \frac{4\chi^2 \kappa}{\kappa^2 + 4\chi^2} \bar{n}, \tag{1}$$

where $\bar{n}$ is the thermal photon number population in the readout resonators, $\kappa$ the resonator linewidth and $\chi$ the qubit's dispersive shift (qubit parameters are listed in Supplementary Table 1).

To measure the effect of the multiplexer on qubit coherence in the presence of well-known slow temporal fluctuations, energy relaxation times $T_1$ and coherence times $T_{2e}$ of three different qubits are continuously measured over approximately 12 hours. All qubit drive and readout signals are routed together either through the multiplexer or through the reference line when the multiplexer is off (Figure 3a).

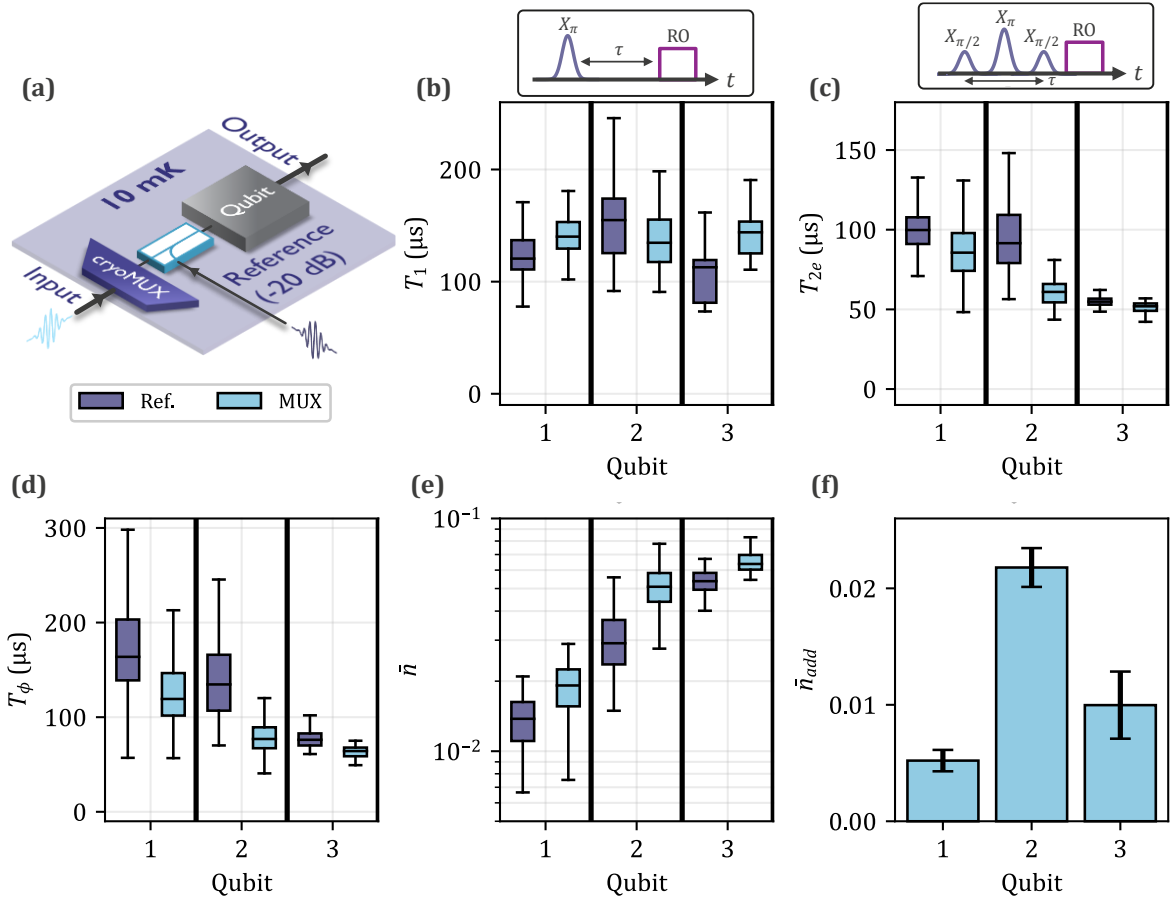

Figure 3: *Superconducting qubit performance with the cryoCMOS multiplexer directly connected to the feedline. (a) Input line multiplexing configuration. A directional coupler is used to obtain reference measurements of qubit coherence times. All drive and readout pulses are routed through the same line ("reference" or "input") for reference ("Ref.") or multiplexed ("MUX") measurements, respectively. (b, c) Comparison of qubit relaxation ($T_1$) and coherence ($T_{2e}$) times between reference and multiplexed measurements (VDD = 0.55 V) for three different qubits (parameters in Supplementary Table 1). Insets: pulse sequences used for $T_1$ and $T_{2e}$ extraction. (d) Dephasing times calculated from $T_1$ and $T_{2e}$. (e) Equivalent total and (f) added photon populations in the readout resonators. Median values are indicated by the horizontal lines in the boxplots; boxes span the interquartile range (Q1-Q3); whiskers represent 1.5x IQR.*

| Qubit | $\Gamma_{\phi,add}$ (kHz) | $\bar{n}_{add}$ (photons/s/Hz) |
|---|---|---|
| 1 | 2.3 ± 0.4 | 0.005 ± 0.001 |
| 2 | 5.6 ± 0.4 | 0.022 ± 0.002 |
| 3 | 2.4 ± 0.7 | 0.009 ± 0.003 |

Table 1: Added dephasing rate caused by the RF MUX on each of the three measured qubits in Figure 3. $\Gamma_{\phi,add} = \Gamma_{\phi,MUX} - \Gamma_{\phi,ref}$. $\bar{n}_{add}$ is obtained using Equation 1. The mean standard error is obtained using $SE = \sqrt{\sigma_{ref}^2/N_{ref} + \sigma_{MUX}^2/N_{MUX}}$ with N the number of datapoints. Qubit parameters are listed in Supplementary Table 1.

Time-averaged $T_1$ times, with median values above 100 μs, show no systematic reduction when measured through the multiplexer (Figure 3b), similarly to what was observed previously[20]. A small degradation was observed in $T_{2e}$ and therefore in $T_\phi$, corresponding to increased dephasing rates of 2 to 5 kHz or added thermal photon populations of approximately 0.005 to 0.02 photons/s/Hz in the readout resonator (Figures 3d-f, Table 1). This difference is found to be statistically significant based on a two-sample Welch t-test[35] ($p \leq 0.013$). Differences in added photon populations between qubits may arise from coherence time measurement uncertainty due to slow fluctuations on hour- or day-long timescales[36–38]. Nevertheless, the additional dephasing remained sufficiently small to allow coherence times of the order of 100 μs to be measured, making the multiplexer suitable for scalable qubit characterization.

While the qubits used in this work do not have dedicated drive lines, the presented results suggest that the multiplexing of XY drive lines would have a negligible effect on $T_1$, similarly to the presented input line multiplexing results. This is because the coupling between a typical XY drive line and a qubit is comparable to the effective coupling between the feedline and the qubit mediated by the readout resonator (detuned by ~2 GHz) in this work (see Supplementary Section 6 for details).

Arguably the most relevant multiplexing is that of qubit readout output lines (Figure 4a), as they contain non-scalable, expensive, centimetre-size microwave components such as circulators, traveling wave parameters amplifiers (TWPA), filters, high-electron mobility transistor (HEMT) amplifiers, and superconducting cables. In the following experiment, the qubit chip's readout output line is directly connected to the RF1 port of the multiplexer, and the RFC port is connected to the standard output line of a dilution refrigerator (See Supplementary Fig. 1).

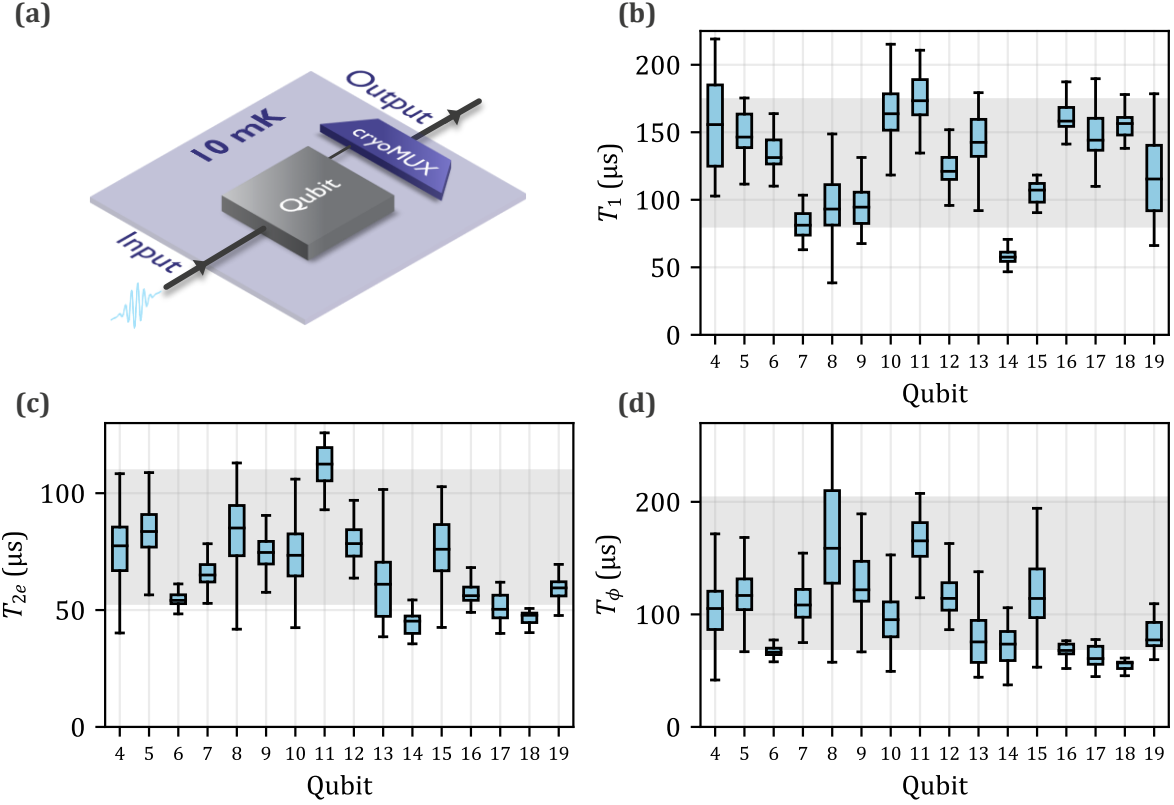

*Figure 4: Superconducting qubit characterization using the cryoCMOS multiplexer on the output line. (a) Output line multiplexing setup. (b-d) Relaxation, decoherence and dephasing times measured through the multiplexed output line. The shaded regions span the interquartile ranges of the reference-line data from Figure 3 and are included here for comparison with multiplexed output line measurements.*

Coherence times measured through the multiplexer at the output show median $T_1$ values of 100-200 μs and $T_{2e}$ values of 50-100 μs, leading to $T_\phi \approx$ 100-200 μs (Figure 4b-d). These values are comparable to those measured using the reference input line (the grey horizontal area in Figure 4 indicates the range of reference values from Fig. 3b-d), and are consistent with coherence times previously reported for similar qubits[32]. This confirms that the multiplexer can be effectively used for output line multiplexing and can accurately capture coherence times up to ~100 μs.

Finally, the cryoCMOS device is used for the multiplexing of low-frequency flux-bias lines, normally used for qubit frequency tuning[15], despite reports of increased 1/f noise in CMOS devices at cryogenic temperatures[39–41]. For this demonstration, the multiplexer was directly connected to the flux line of a superconducting qubit chip and to a dedicated low-frequency signal input line (Figure 5a).

To assess the impact of the multiplexer's noise on the flux line, including low-frequency (1/f) noise, a standard flux noise spectroscopy protocol[42,43] is implemented using a frequency-tuneable transmon qubit containing a magnetic flux-sensitive superconducting interference device (SQUID, Figure 5b). In this spectroscopy protocol, the qubit's $T_1$ and $T_{2e}$ are measured as a function of statically applied flux bias ($\Phi_e$, Figure 5c) and the dephasing rate associated with the Hahn-echo dephasing time ($\Gamma_\phi^e = 1/T_{2e}$, Figure 5d, e) is extracted as a function of the flux dispersion of the qubit's ground to excited state transition frequency $\partial f_q/\partial \Phi_e$. The multiplexer remains powered on throughout the measurement. In Figures 5d and 5e, only the flux-sensitive part of the dephasing is shown, by subtracting the intrinsic dephasing measured at the sweet spot, i.e. $\Gamma_\phi^e - \Gamma_{\phi,SS}^e$.

We model the qubit's sensitivity to flux noise using a power spectral density (PSD) comprising of a 1/f component and a white component: $S_\phi(\omega) = \frac{A}{|\omega|} + B$ (see Supplementary Section 8). The experimentally extracted dephasing rates values are fitted to the following dephasing model to extract the noise parameters $A$ and $B$:

$$\Gamma_\phi^e \approx \sqrt{A \ln(2)} \left|\frac{\partial \omega_q}{\partial \phi_e}\right| + B\pi \left(\frac{\partial \omega_q}{\partial \phi_e}\right)^2, \quad (2)$$

where $\omega_q = 2\pi f_q$. The $A$ and $B$ parameters are rescaled to physical units of magnetic flux ($\Phi$) and ordinary frequency ($f$) in Figure 5.

In this experiment, we use flux-dependent $T_{2e}$ (Hahn echo) pulse sequences instead of Ramsey $T_2^*$ sequence to quantify the flux noise parameters in order to minimize effects from strong intrinsic dephasing (e.g. from slowly varying frequency drifts due to TLS defects[36,44] or charge parity jumps typically observed in transmon qubits with long coherence times[45]). While the Hahn echo sequence effectively acts as a high-pass filter with a cut-off frequency of a few 10 kHz[43], it is still sensitive to part of the 1/f noise spectrum. The derivation of Equation 2 takes the echo filter function into account[43].

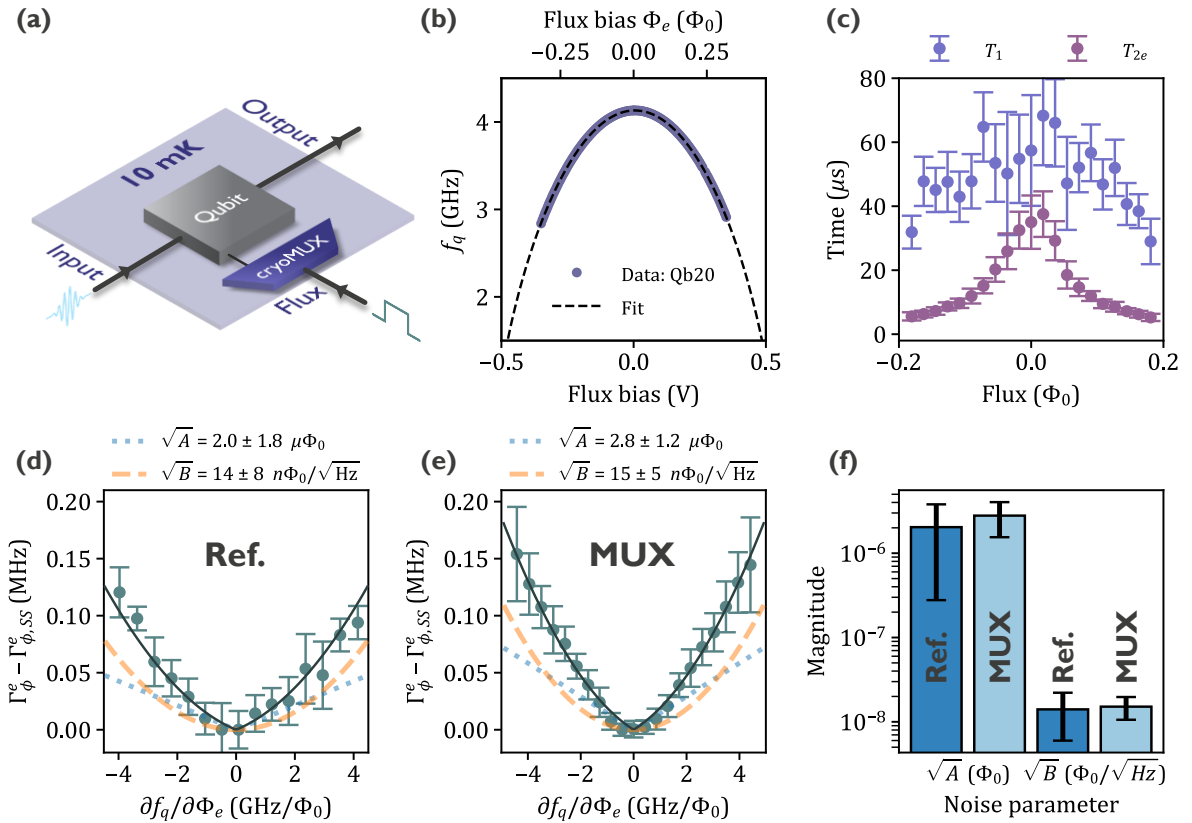

Figure 5: Flux-dependent coherence time measurements used to determine flux-noise parameters. (a) Experimental setup: the flux line of a flux-tuneable superconducting qubit is directly connected to a multiplexer. (b) Qubit frequency $f_q$ as a function of applied flux bias. The data is fitted to a square root-cosine model. (c) Qubit $T_1$ and $T_{2e}$ as a function of flux bias applied through the multiplexer. (d, e) Qubit dephasing rates extracted from Hahn-echo experiments for reference and multiplexed measurements, respectively. Values are offset by the dephasing rate at the sweet spot. The reference measurements are extracted from another, identical qubit measured in a separate cooldown without the multiplexer. (f) Comparison of the extracted noise parameters between reference and multiplexed configurations.

The 1/f noise term ($\sqrt{A} \approx 2.8\,\mu\Phi_0$) and the white noise parameters ($\sqrt{B} \approx 15\,\text{n}\Phi_0/\sqrt{\text{Hz}}$, equivalent to a current noise spectral density of ~ 2.5 x $10^{-22}$ A²/Hz on the flux line) are compared to measurements extracted from an identical qubit located on another chip and measured during a separate cooldown without the multiplexer (labelled *"Ref."* in Figure 5). The noise parameters do not significantly increase during multiplexed operation within measurement uncertainty (Figure 5f). The obtained *A* and *B* parameters are also comparable to previously published data[43] when extracted with our model (Supplementary Section 8). These results indicate that the flux noise added by the multiplexer is not notably larger than the intrinsic flux noise originating from electromagnetic noise in the low frequency lines as well as from surface spins[46] and quasiparticle tunnelling[47].

At the largest static flux-bias used in the experiment ($\pm 0.2\,\Phi_0$), a DC current of 0.23 mA flowing through the multiplexer results in approximately 0.28 µW of Joule heating in the series RF transistor, based on the simulated on-state resistance resistance $R_\text{on}$ of ~5.3 Ω. This heating increases the mixing chamber temperature by up to 5 mK, resulting in a $T_1$ reduction of ~30% (Figure 5c). Despite this, the multiplexer still enables a qubit frequency tuning range exceeding 1 GHz, sufficient for the characterization of flux-line–SQUID mutual inductances, or for performing TLS spectroscopy[36,44]. The power dissipation can be further reduced by using short flux pulses rather than static currents, and by increasing multiplexer's VDD (thereby reducing $R_\text{on}$) at the cost of increased static power consumption by the cryoCMOS circuit itself (See Supplementary Figure 5 and 6).

## IV. Discussion

The results presented in this work demonstrate that cryoCMOS can be used for multiplexing all signal lines in superconducting qubits, including readout input and output lines, flux bias lines and, in principle, XY drive lines, with low or negligible impact on qubit coherence. This makes a cryoCMOS multiplexer an attractive component for scalable qubit characterization, which is urgently needed for the characterization of a large number of quantum devices fabricated at an industrial scale[32,48]. In addition to testing quantum device functionality, large-scale characterization is also valuable for materials and fabrication process exploration, as well as for any other cryogenic experiment requiring DC and RF signals up to 10 GHz.

The extremely low static power of 200 pW allows scaling the characterization system to as many as 100,000 multiplexers operating in static mode and addressing up to 400,000 devices (assuming 20 µW cooling power at the mixing chamber stage and excluding flux bias multiplexing) which is far greater than the physical capacity of today's dilution refrigerators. The number of addressable ports can be further increased by designing multiplexers with more than four ports. Doing so increases the total power consumption only by the additional shunt transistor gate leakage contribution, which is approximately 25% of a single large series RF transistor.

Reducing the remaining power consumption is essential for developing more complex cryoCMOS circuits operating at millikelvin temperatures in close proximity to quantum devices. As shown in this work, the dominant static power contribution originates from gate leakage through the large RF switch transistors' gates. At GHz frequencies, gate leakage can generate electromagnetic noise through two possible mechanisms: (i) shot noise from random fluctuations in the tunnelling gate current[49–51] or (ii) thermal noise from self-heating[52]. Both mechanisms could be mitigated by lowering the chip's supply voltage (VDD) using lower-$V_T$ transistors (such as *slvt* type transistors in the digital circuit), tuning the transistor's threshold voltages using FDSOI-specific back-gate biasing, or employing thicker gate oxides. Thermal noise could be further reduced by improving heat flow and thermalization within the chip and to the environment. Although thermalization studies at cryogenic temperatures down to 4K are underway[52,53], a detailed understanding of thermal processes at millikelvin temperatures is still lacking and would significantly advance the design and optimization of cryoCMOS devices operating in this regime.

## V. Conclusion

We have presented an ultra-low power cryoCMOS RF multiplexer fabricated in the FDSOI technology and demonstrated its compatibility with superconducting qubits at millikelvin temperatures for the multiplexing of all the signal types required for superconducting qubit operation. The device showed a record-low static power dissipation of 200 pW and RF performance compatible with qubit operation across the 0-10 GHz frequency band. We have demonstrated both input and output readout line multiplexing with the cryoCMOS chip directly connected to qubits. In addition, we showed for the first time that the multiplexer's 1/f noise, relevant for flux-bias line multiplexing, is not higher than the intrinsic 1/f noise seen by the qubit, thereby enabling flux line multiplexing. The presented achievements can be directly applied to scalable quantum device characterization and testing and mark a significant milestone towards the development of CMOS controlled scalable quantum computing architectures at millikelvin temperatures.

## VI. Author contributions

## VII. Acknowledgements

The authors gratefully thank Michael Libois for support with PCB design and Koen Verhemeldonck for wire-bonding support. This work is supported in part by the imec Industrial Affiliation Program on Quantum Computing. L.F. acknowledges the support of the Research Foundation - Flanders through the Strategic Basic Research PhD program (grant no. 1SA0426N). K.D.G. acknowledges support from Proximus as 600-year KU Leuven Proximus Chair in Quantum Science and Technology. This work is supported by the Chips JU project ARCTIC (Project #101139908). That project is supported by the Chips Joint Undertaking and its members (including top-up funding by Belgium, Austria, Germany, Estonia, Finland, France, Ireland, The Netherlands and Sweden). ARCTIC gratefully acknowledges the support of the Canadian and the Swiss federal governments.

# Supplementary Information

## 1. Methods

**Experimental setup**

All qubit and cryoCMOS cryogenic measurements were performed at 10mK inside a Bluefors LD400 dilution refrigerator with standard wiring to the superconducting qubits (Supplementary Figure 1).

The cryoCMOS multiplexer chip is wire-bonded to a four-layer PCB with Rogers RO4350B low-loss dielectric (Figure 1b) and is thermally anchored to the mixing plate (10 mK stage). The chip is supplied by a Keysight B2901B source measure unit (SMU) fitted with the Keysight N1297A addon for triaxial outputs, enabling the accurate measurement of low currents (< 1nA). Port selection (RF 1-4, pins D0, D1) is done using 2 AWG (Keysight M3202A) channels, which are filtered using low-pass filters (Minicircuits VLFX-780+) and programmed according to the binary number represented by (D1 D0) + 1, with D0 as the least significant bit.

Static power consumption is measured by sweeping the supply voltage while recording the supply current. Two separate supply ports are available on the chip: one supplying the core digital and RF circuits and the other supplying the ESD protection cells. In normal operation, both pins are connected to the supply port of the SMU. When exploring different supply voltages for the core and ESD circuits, an additional low-noise voltage source (Bilt iTest BE2142) is used. Dynamic power is determined from supply voltage and current measurements while switching D0 using a square wave signal at various frequencies (1 kHz – 1 MHz).

The RF properties of the chip (insertion loss, isolation) are measured in two separate fridge cooldowns. During the first cooldown, the insertion loss and isolation are measured through the multiplexer. In a subsequent cooldown, a calibration measurement is obtained by bypassing the multiplexer and directly connecting the input and output lines at the 10 mK stage. Scattering parameters are measured using a Keysight P5004B vector network analyser.

Qubit samples (Supplementary Figure 4) are wire-bonded into a high-purity aluminium sample holder and fastened to a high-purity oxygen-free copper vertical mount, thermally anchored to the mixing chamber. To protect the samples from stray magnetic fields and infrared radiation, they are enclosed in a cryoPerm shield and a copper radiation shield at the millikelvin stage, as well as in a standard μ-metal shield mounted in the vacuum can of the dilution refrigerator.

**Simulations**

Circuit simulations are performed in Spectre using the foundry PDK at -40°C, which is the lowest temperature covered by the foundry models. These simulations served as a proxy for cryogenic behaviour in the absence of a full cryogenic PDK. The S-parameters of the on-die transmission lines, including wire-bonds, are simulated using a Maxwell EM solver.

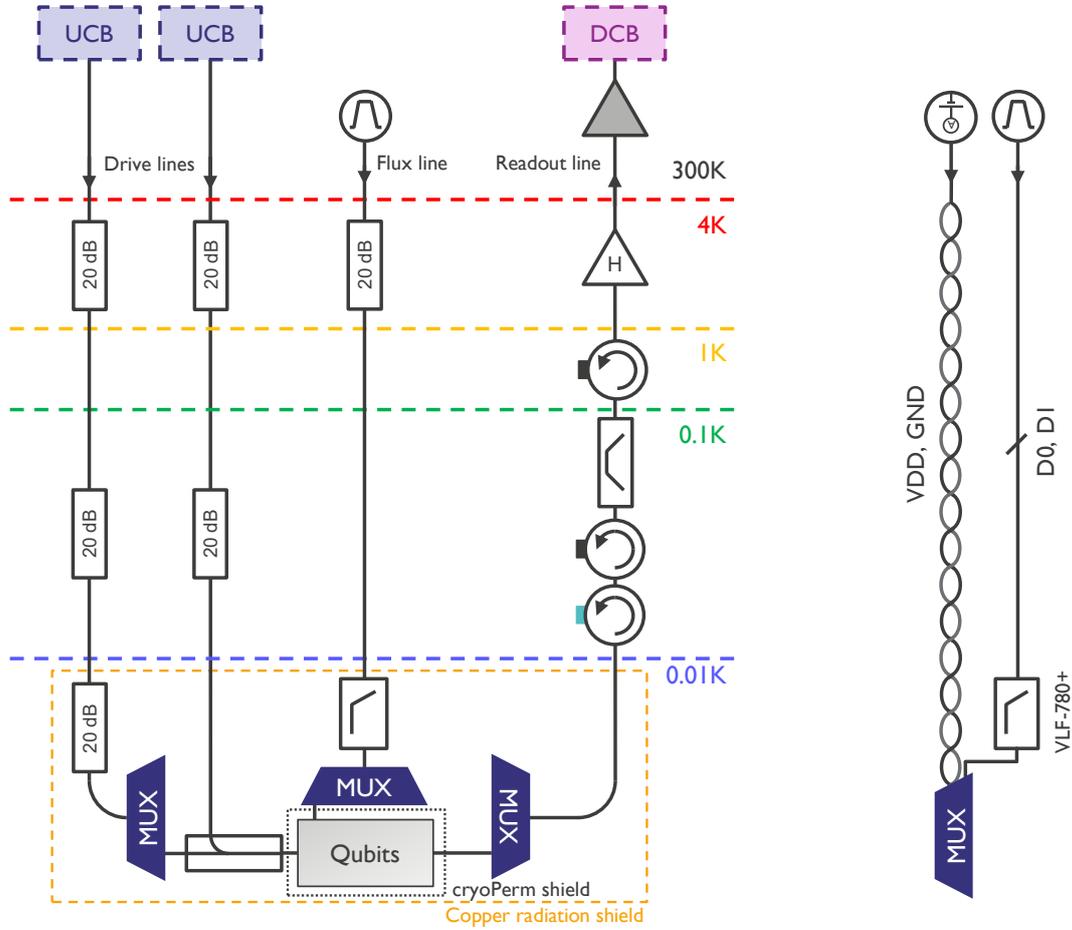

Supplementary Figure 1: Room temperature and cryogenic experimental setup for qubit characterization using the cryoCMOS multiplexer. While all three multiplexers are shown connected in the diagram, only one multiplexer was used at a time.

## 2. Power consumption of the different circuit elements

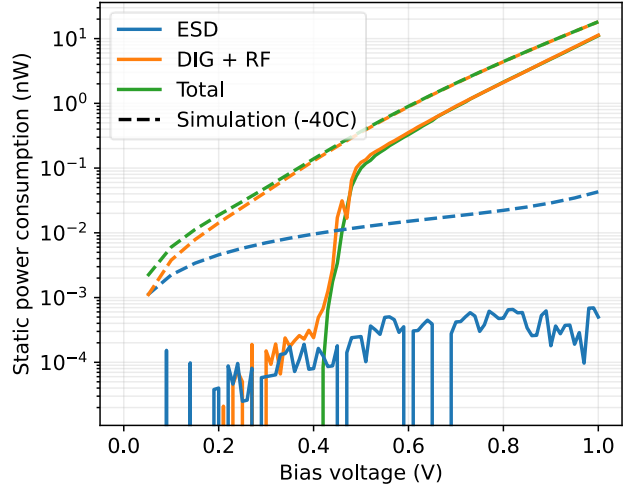

*Supplementary Figure 2: Static power consumption of the different parts of the multiplexing chip. The current supplied to the electrostatic discharge (ESD) protection cells is below the measurement noise floor of the SMU (< 100 fA). Solid lines correspond to measurements performed at 10 mK, while dashed lines show simulations at -40°C.*

## 3. Dynamic power consumption

The dynamic power consumption of the chip increases linearly with switching frequency for each supply voltage. From the corresponding slopes, the energy dissipated per switching event is extracted, considering that a switching frequency of 1 Hz (square wave frequency on the AWG) corresponds to 2 switching events per second. The extracted slopes follow a quadratic dependence, consistent with the capacitive power-loss formula $P_{dyn} = fCV^2$, where the effective capacitance includes contributions from the transistor gate capacitances as well as other parasitic capacitances in the chip. The dissipation per voltage squared (0.715 pW/Hz/V$^2$) is comparable to a previously published cryoCMOS multiplexer (~ 1 pW/Hz/V$^2$)[20], though the absolute dissipation is lower due to operation at a reduced supply voltage.

Circuit simulations at -40°C predict an increased dynamic power dissipation compared to 10 mK measurements. Since the MOSFET gate capacitance in inversion is only weakly temperature-dependent, this discrepancy cannot be attributed to changes in $C$. It could however be explained by variations in circuit switching dynamics. At cryogenic temperatures, transistors are capable of a larger current drive than at room-temperature, which enables the small digital control gates to switch the large RF transistors more rapidly, reducing the crossover duration when both NMOS and PMOS are in their conductive mode. This results in a smaller shoot-through current dissipation per switching event.

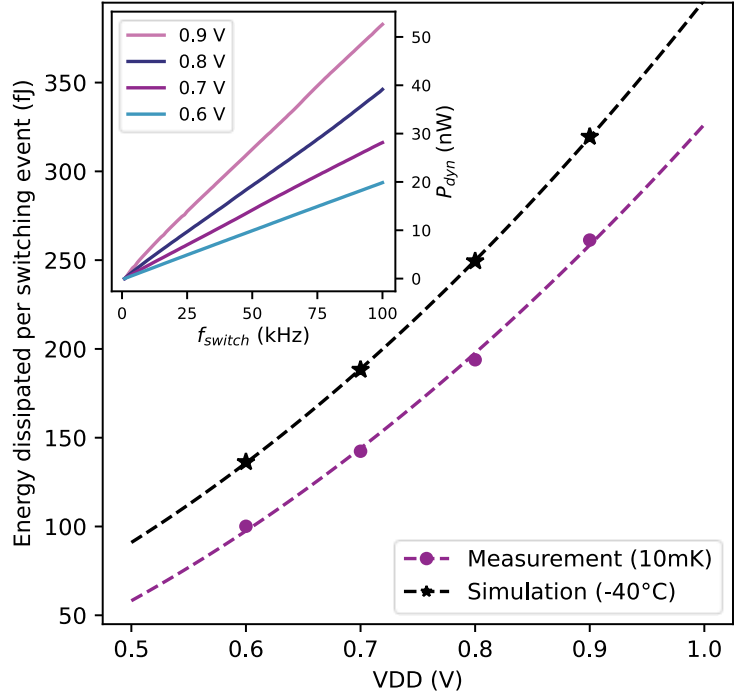

*Supplementary Figure 3: Dynamic power consumption of the multiplexing chip. The points represent the energy dissipated during each switching event, with a quadratic fit (dashed line) serving as a guide to the eye. The black stars represent simulation results at -40 °C temperature. Inset: dynamic power as a function of switching frequency, for various supply voltages.*

## 4. Qubit samples

The qubits used in this work to characterize the multiplexer are fabricated in a 300mm foundry-compatible cleanroom at imec[32]. Each chip contains 10 transmon qubits of different sizes, all coupled to a common feedline via readout resonators (Supplementary Figure 4a). Only one qubit on the chip is frequency-tuneable (Supplementary Figure 4b). Its smaller capacitor size and higher participation ratio leads to shorter relaxation times ($T_1$) than the other qubits on the chip[32].

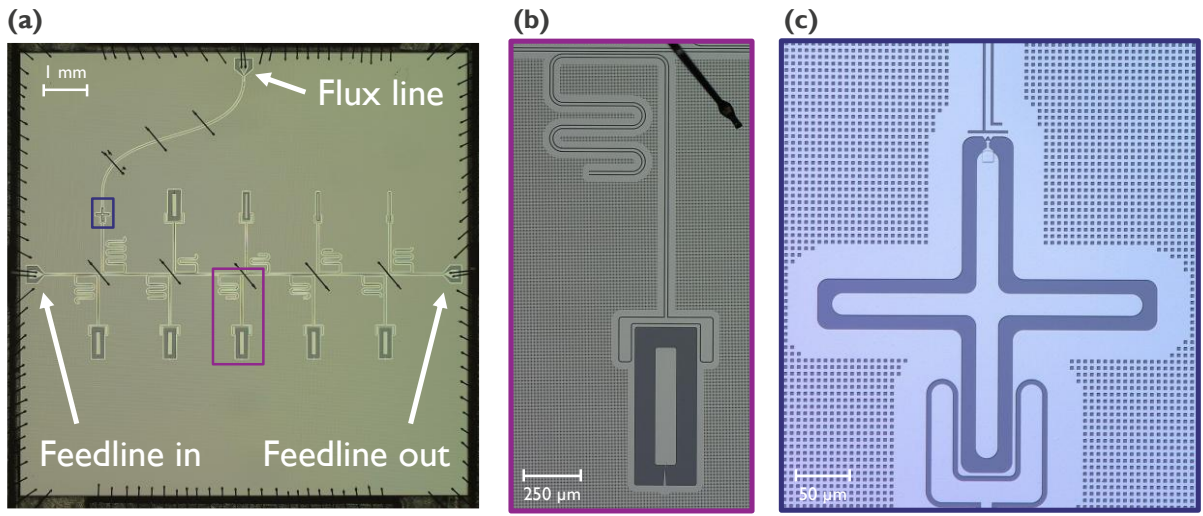

*Supplementary Figure 4: (a) Micrograph of a packaged transmon qubit chip used for multiplexer characterization. Each chip contains 10 qubits, including one frequency-tuneable qubit. (b) Zoom-in on a fixed-frequency qubit. (c) Zoom-in on the frequency-tuneable qubit and its flux line.*

The following qubit parameters were measured using standard continuous-wave or pulsed experiments for superconducting resonator and qubit characterization. These parameters are used for calculating the resonator photon populations in the main text.

| Qubit | $f_q$ (GHz) | $f_r$ (GHz) | $\kappa/2\pi$ (MHz) | $\chi/2\pi$ (MHz) | $\overline{T}_1$ (µs) | $\overline{T}_{2e}$ (µs) | $\overline{T}_\phi$ (µs) | $\overline{n}_{th}$ (µs) |
|---|---|---|---|---|---|---|---|---|
| 1 | 3.514 | 7.831 | 0.709 | - 0.117 | 126 | 105 | 160 | 0.014 |
| 2 | 3.292 | 6.615 | 1.064 | - 0.106 | 152 | 93 | 129 | 0.03 |
| 3 | 3.554 | 7.018 | 0.988 | - 0.100 | 106 | 55 | 76 | 0.054 |
| 4 | 2.987 | 6.595 | - | - | 158 | 77 | 96 | - |
| 5 | 3.039 | 6.797 | - | - | 147 | 84 | 116 | - |
| 6 | 3.157 | 6.998 | - | - | 136 | 55 | 67 | - |
| 7 | 3.489 | 7.377 | - | - | 82 | 66 | 109 | - |
| 8 | 3.552 | 7.824 | - | - | 95 | 84 | 156 | - |
| 9 | 3.607 | 6.501 | - | - | 94 | 74 | 122 | - |
| 10 | 3.666 | 6.705 | - | - | 165 | 75 | 91 | - |
| 11 | 3.596 | 7.115 | - | - | 174 | 111 | 163 | - |
| 12 | 4.001 | 6.503 | - | - | 123 | 80 | 111 | - |
| 13 | 3.793 | 6.707 | - | - | 143 | 61 | 72 | - |
| 14 | 3.893 | 7.373 | - | - | 58 | 44 | 70 | - |
| 15 | 3.861 | 7.821 | - | - | 107 | 84 | 118 | - |
| 16 | 3.989 | 6.499 | - | - | 161 | 57 | 68 | - |
| 17 | 3.881 | 6.703 | - | - | 148 | 52 | 62 | - |
| 18 | 4.020 | 6.908 | - | - | 155 | 47 | 55 | - |
| 19 | 3.984 | 7.112 | - | - | 119 | 59 | 80 | - |
| 20 | 4.137 | 6.055 | 0.423 | - 0.253 | 57 | 35 | 50 | 0.013 |

*Supplementary Table 1: Parameters of the qubits used in this work. A bar above the parameter symbol represents mean values. The values with dashes (-) were not measured.*

## 5. Statistical analysis

To quantitatively assess whether the multiplexer introduces statistically significant changes in qubit coherence times, a statistical analysis of coherence time measurements was performed for the reference line and the multiplexer line. For each qubit, a series of interleaved energy relaxation time ($T_1$) and echo coherence time ($T_{2e}$) measurements were acquired over several hours.

The Welch t-test[35] was used to test the null hypothesis $H_0$ that the multiplexer does not introduce any statistically significant noise to the qubit's environment, i.e. that the two independent datasets share the same mean:

$$H_0: \mu_{REF} = \mu_{MUX}, H_1: \mu_{REF} \neq \mu_{MUX}. \qquad (S1)$$

The analysis was carried out using a Welch two-sample t-test (computed in Python using SciPy[54]), assuming unequal variances. A value of $p < 0.05$ is considered statistically significant (validating $H_1$). The resulting t- and p-values are presented in Supplementary Table 2.

All qubits show small but statistically significant differences in $T_{2e}$ and in the corresponding $\Gamma_\phi$ (with $p \ll 0.05$). $T_1$ values also exhibit small p values (Qb1, Qb3), which could suggest a statistically significant difference, but they were dismissed as the sign of the difference indicates increased relaxation times in the multiplexed case, contrary to what can be physically expected.

| Value | Qb1 Welch t | p | Qb2 Welch t | p | Qb3 Welch t | p |
|---|---|---|---|---|---|---|
| $T_1$ | -3.89 | 0.0002 | 1.78 | 0.0778 | -3.7 | 0.0018 |
| $T_{2e}$ | 3.28 | 0.0013 | 11.9 | $\ll 0.0001$ | 2.72 | 0.0124 |
| $\Gamma_\phi$ | 4.98 | $\ll 0.0001$ | 12.79 | $\ll 0.0001$ | 3.97 | 0.0008 |

*Supplementary Table 2: Results of the Welch t-test on the data shown in Figure 3. Comparisons between the reference and multiplexed datasets for each qubit show low p-values, confirming hypothesis $H_1: \mu_{REF} \neq \mu_{MUX}$.*

## 6. Multiplexer impact on XY drive line

To assess the impact of the multiplexer on the qubit when connected to a XY drive line, the typical coupling rate between the line and a transmon qubit,

$$\kappa_d = \frac{C_d^2}{C_\Sigma} Z_d \omega_q^2, \quad (S2)$$

is compared to the qubit-feedline coupling mediated by the detuned readout resonator (Purcell decay rate) in our experiment,

$$\gamma_P \approx \frac{g^2 \kappa}{\Delta^2}, \quad (S3)$$

with $g$ the qubit-resonator coupling, $\kappa$ the resonator-feedline and $\Delta$ the detuning between the qubit and resonator frequencies[19]. Using the typical transmon parameters listed in Supplementary Table 3, the corresponding rates are found to be $\kappa_d = 2\pi \times 0.31$ kHz and $\gamma_P = 2\pi \times 0.586$ kHz, confirming that noise from the multiplexer on a dedicated XY drive line would have a similarly negligible effect on qubit energy relaxation rates as reported in Fig. 3b.

| Parameter | Symbol | Value |
|---|---|---|
| Qubit frequency | $\omega_q$ | 3.5 GHz |
| Resonator frequency | $\omega_r$ | 7 GHz |
| Resonator loaded quality factor | $Q_l$ | 5000 |
| Drive line-transmon capacitance | $C_d$ | 0.1 fF |
| Total transmon capacitance | $C_\Sigma$ | 105 fF |
| Drive line impedance | $Z_d$ | 50 Ω |

*Supplementary Table 3: Typical qubit parameters used to calculate coupling rate estimates.*

## 7. Multiplexer impact on flux bias line

The impact of flux line multiplexing on qubit energy relaxation times is also investigated when varying the multiplexer's supply voltage. For VDD values of up to 1 V and with no DC current through the multiplexer, no measurable impact on the qubit's coherence times is observed compared to a reference measurement in which the multiplexer is turned off. This result indicates that the noise added to the flux line by the multiplexer has a negligible effect on coherence times.

When the multiplexer is turned on (VDD = 0.55V) and the qubit is biased to 0.2 $\Phi_0$ (corresponding to a 0.23 mA static current through the multiplexer), the qubit's $T_1$ is reduced by approximately 30% compared to its value at the sweet spot. This reduction is attributed to increased thermal noise generated by Joule heating inside the large RF transistor, as discussed in the main text.

Interestingly, the qubit's $T_1$ is recovered when the supply voltage is increased from 0.55V to 1 V. This is most likely due to the reduced on-state resistance of the series RF transistor at higher supply voltages, and with that reduced Joule heating (Supplementary Figure 6). For efficient flux bias multiplexing, an optimal VDD can be chosen that minimizes Joule heating due to the DC current, at the expense of increased static power dissipation.

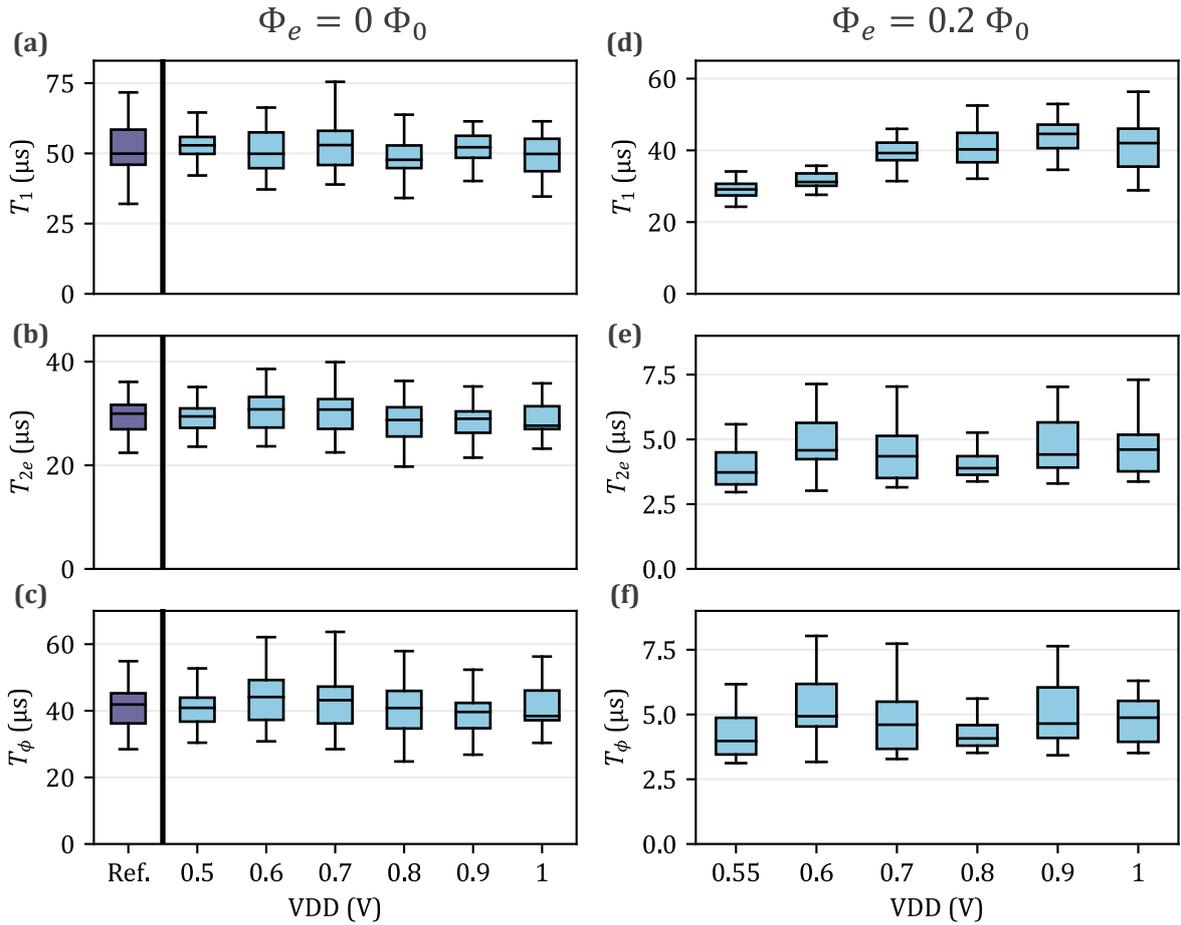

*Supplementary Figure 5: Qubit lifetimes with multiplexed flux line, as a function of multiplexer supply voltage. The times are measured at the sweet spot ($\Phi_e = 0\Phi_0$, a-c) and away from the sweet spot ($\Phi_e = 0.2\ \Phi_0$, d-f).*

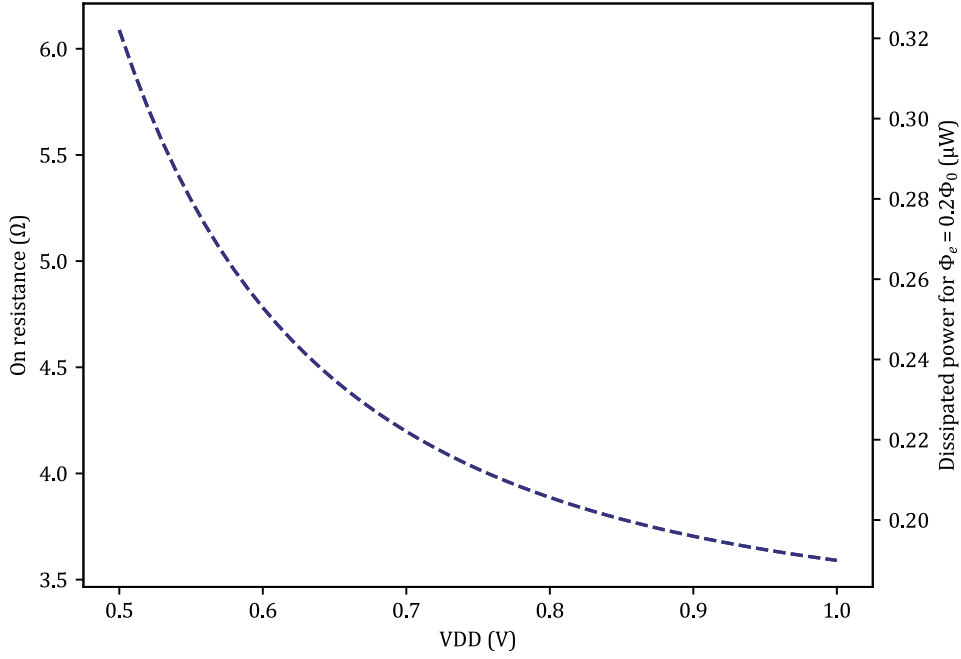

*Supplementary Figure 6: Simulated on-state resistance $R_{on}$ of the series RF transistor at -40 °C (left axis) as a function of the supply voltage VDD. The corresponding dissipated power is calculated for a 0.23 mA DC current (right axis) required to generate an applied flux $\Phi_e = 0.2\Phi_0$.*

## 8. Flux-noise induced dephasing

In the following, Equation 2 from the main text is derived for a Hahn-echo pulse sequence. The derivation is based on the approach presented by Mergenthaler et. al.[43], although modified to accommodate more widely used standard definitions.

Dephasing arises from fluctuations in the relative phase $\phi$ of a coherent superposition state such as $\frac{|0\rangle + e^{i\phi}|1\rangle}{\sqrt{2}}$. The accumulated phase over an interrogation interval is given by the integral of the qubit transition frequency $\omega_q$[19]:

$$\phi(t) = \int_0^\tau \omega_q(t)dt = \langle \omega_q \rangle \tau + \delta\phi(\tau), \tag{S4}$$

Where $\delta\phi(\tau)$ captures the contribution from adiabatic fluctuations of a fluctuator $\lambda$ (e.g., magnetic flux).

$$\delta\phi(\tau) = \frac{\partial \omega_q}{\partial \lambda} \int_0^\tau \partial\lambda(t)dt. \tag{S5}$$

When many weakly coupled fluctuators $\lambda$ contribute to $\phi(t)$, the central limit theorem states that $\delta\phi(\tau)$ is Gaussian-distributed. The phase factor will decay according to:

$$\langle e^{i\delta\phi(\tau)} \rangle = e^{-\frac{\langle \phi^2(\tau) \rangle}{2}} = \exp(-\chi(\tau)), \tag{S6}$$

Where the function $\chi(\tau)$ is

$$\chi(\tau) = \frac{1}{2}\left(\frac{\partial \omega_q}{\partial \lambda}\right)^2 \int_{-\infty}^{\infty} |F(\omega,\tau)|^2 S_\lambda(\omega)d\omega, \tag{S7}$$

here the noise spectral density $S(\omega)$ is filtered by the interrogation function $F(\omega,\tau) = i\tau \sin\frac{\omega\tau}{4} \operatorname{sinc}\frac{\omega\tau}{4}$ for the Hahn-echo pulse sequence (used to extract $T_{2e}$).

## Noise model

We consider the effective flux noise seen by the qubit as the sum of *1/f* and white noise components:

$$S_\lambda(\omega) = \frac{A}{|\omega|} + B. \tag{S8}$$

Substituting the filter function and performing the variable change $\theta = \omega\tau$, $d\omega = d\theta/\tau$, the dephasing function becomes:

$$\chi(\tau) = 8AD^2 \int \frac{\sin^4(\theta/4)}{|\theta|^3} \tau^2 d\theta + 8BD^2 \int \frac{\sin^4(\theta/4)}{\theta^2} \tau d\theta. \tag{S9}$$

Evaluating the integrals gives:

$$\chi(\tau) = AD^2 \ln(2)\, \tau^2 + BD^2 \pi \tau, \tag{S10}$$

where $D = \partial\omega_q/\partial\lambda$. Substituting this expression into Equation S6 shows that dephasing due to *1/f* noise produces a Gaussian decay function ($e^{-(\Gamma_\phi\tau)^2}$), whereas white noise leads to a simple exponential decay ($e^{-\Gamma_\phi\tau}$). As experimentally measured coherence traces are typically well described by a single exponential, we employ the approximation $\exp(-at^2) \approx \exp(-\sqrt{a}\,at)$ to extract the effective exponential dephasing rate. The resulting coherence envelope then becomes:

$$\exp(-\Gamma_\phi\tau) = \exp(-\chi(\tau)) \approx \exp\left(-\left[\sqrt{A_{\phi,\omega}\ln(2)}\,D + B_{\phi,\omega}\pi D^2\right]\tau\right), \tag{S11}$$

which can be directly compared to the dephasing rate $\Gamma_\phi$ extracted from measurements as a function of applied flux (Equation 2 in the main text).

## Rescaling to flux and frequency units

Note that the noise parameters A and B in the derivation above are expressed in terms of phase ($\phi$) and angular frequency ($\omega$). In practice, noise is reported in physical units, in terms of flux ($\Phi$) and frequency ($f$). Using the definitions for flux $\phi = \pi\Phi_e/\Phi_0$, the flux noise spectral density $S_\phi(\omega) = \int \langle\phi(\tau)\phi(0)\rangle e^{-i\omega\tau} d\tau$ and angular frequency $\omega = 2\pi f$, the noise parameters are rescaled as:

$$A_{\Phi,f} = A_{\phi,\omega}/\pi^2 \qquad [\Phi_0^2] \tag{S12a}$$
$$B_{\Phi,f} = 2B_{\phi,\omega}/\pi \qquad [\Phi_0^2/\text{Hz}] \tag{S12b}$$

All values for $A$ and $B$ reported in the main text are expressed in $\Phi$ and $f$.

## Comparison with previously published data

By extracting the noise parameters from the data presented in Figure 2 of ref. 43 using Eq. S11, we obtain $\sqrt{A_{\Phi,f}} = 0.5~\mu\Phi_0$ and $\sqrt{B_{\Phi,f}} = 7.4~\text{n}\Phi_0/\sqrt{\text{Hz}}$, which is comparable to those obtained in the present work (Figure 5d, e).